\documentstyle[12pt]{article}
\begin{document}
\begin{flushright}
IHEP 96-83\\
hep-ph/9610482
\end{flushright}
\begin{center}
{\large\bf Estimate of $\alpha_s(m_Z)$ in sum rules for bottomonium}\\
\vspace*{3mm}
V.V.~Kiselev \\
\vspace*{3mm}
Institute for High Energy Physics,\\
Protvino, Moscow Region, 142284, Russia\\
E-mail: kisselv@vxcern.cern.ch
\end{center}

\begin{abstract}
The value of $\alpha_s(m_Z)=0.106\pm 0.005$ is obtained in
a scheme of sum rules for the leptonic constants of the vector $nS$-levels
in the heavy $(\bar b b)$-quarkonium.
\end{abstract}

\section{Introduction}

The coupling constant of strong interaction measured in the $Z$-boson 
pole at LEP in the framework of Standard Model, $\alpha_s(m_Z)\approx 0.125$
is significantly greater than the value of $\alpha_s(m_Z)\approx 0.11$,
which is extracted from the low-energy data \cite{1}, such as the
deep inelastic lepton-hadron scattering (the Gross--Llewellyn Smith sum rule
and the $Q^2$ evolution of structure functions \cite{kat}), 
the $\Upsilon$ decays,
in the QCD sum rules for the $(\bar b b)$ system, and in the lattice 
computations of mass spectrum for the heavy quarkonium. The mentioned deviation
between the LEP measurements at $Q^2\sim 10^4$ GeV$^2$ and the value,
expected from QCD at lower energies ($Q^2\sim 10^2$ GeV$^2$), can point out
virtual effects of a "new" large-energy scale physics, which is beyond
the scheme of Standard Model, at the energies of $Z$-boson \cite{2}.

The QCD sum rules for the leptonic constants of vector states in the
heavy $(\bar b b)$-quarkonium have been recently considered for the 
precision extraction of the strong-interaction coupling constant and the
$b$-quark mass in \cite{3}. M.Voloshin has used the scheme of moments for the
spectral density of the transversal correlator of vector currents in the
nonrelativistic approximation, so that he has found 
$\alpha_s(m_Z)=0.109\pm 0.001$, where the statistical error is presented only.
To study a methodic stability of the result on the $\alpha_s(m_Z)$
measurement in the sum rules for $(\bar b b)$, a consideration of the
problem in other schemes is of a great interest.

In this paper we find $\alpha_s(m_Z)=0.106\pm 0.005$ from the data on the
constants of vector $nS$-levels of $(\bar b b)$ in the sum rule scheme
offered in \cite{4}.

\section{Evaluation of $\alpha_s$}

In \cite{4} one has considered the scheme of sum rules \cite{5} for the
two-point transversal correlator of vector currents of heavy quarks. It
allows one to use the nonrelativistic movement of heavy quarks inside the
quarkonium, the suppression of nonperturbative corrections from the
quark-gluon condensates in the expansion over the inverse heavy-quark mass
as well as the regularity of the heavy quarkonium mass spectra and
explicit expressions for the energetic density of $nS$-levels. One has found
the broad region of numbers $n_{mom}$ for the moments of the spectral density, 
where the following conditions are valid: the contribution of gluon 
condensate is low and the terms from the excited resonances in the heavy
quarkonium are still essential (for $(\bar b b)$ one has $2< n_{mom}<20$).
In this region the moments from the contribution of the sum over resonances
can be quite accurately represented in the integral form with the energetic
density of levels $dn/dM_n$, where $n$ is the number of $nS$-level in the
quarkonium. So, the following relation for the leptonic constants
of vector states takes place
\begin{equation}
\frac{f_n^2}{M_n} = \frac{\alpha_s}{\pi}\; \frac{dM_n}{dn}\;
\biggl(\frac{4m_{red}}{M_n}\biggr)^2 H_V\; Z_{\rm sys}\;,
\label{1}
\end{equation}
where $m_{red}=m_1m_2/(m_1+m_2)$ is the reduced mass in the system of 
two quarks with the masses $m_{1,2}$. In the $\overline{\rm MS}$ 
renormalization scheme the $\alpha_s$ value is determined by the expressions
\begin{equation}
\alpha_s= \alpha_s^{\overline{\rm MS}}(e^{-5/3} \bar p_Q^2)\;,
\label{2}
\end{equation}
where $\bar p_Q^2$ is the average square of the momentum transfer between 
the quarks, so
\begin{equation}
\bar p_Q^2 = \langle ({\bf p}_1 -{\bf p}_2)^2\rangle =
2\langle {\bf p}^2_{1,2}\rangle = 4\langle T\rangle \; m_{red}\;.
\label{3}
\end{equation}
The $H_V$ factor is the result of the hard gluon correction to the 
correlator of vector currents. It takes the form
\begin{equation}
H_V = 1+\frac{2\alpha_s^H}{\pi}\biggl(\frac{m_2-m_1}{m_2+m_1}\ln\frac{m_2}{m_1}
-\frac{8}{3}\biggr)\;,
\label{4}
\end{equation}
where for $m_1=m_2=m_Q$ the QCD coupling constant according to the BLM
procedure \cite{blm} is determined by the expression \cite{3}
\begin{equation}
\alpha_s^H = \alpha_s^{\overline{\rm MS}}(e^{-11/12}m_Q^2)\;.
\label{5}
\end{equation}
Further, the factor $Z_{\rm sys}=Z_{\rm nr}/Z_{\rm int}$ is close to
unit and it determines the systematic correction due to the nonrelativistic
approximation for the contribution of quark loop with the account for the
$\alpha_s/v$-terms of the coulomb type ($Z_{\rm nr}$) and due to the
integral representation of the resonance contributions into the hadronic
part of sum rules ($Z_{\rm int}$). 
The $Z_{\rm sys}$ value weakly depends on the number of
$nS$-resonance, so that for the basic state of $(\bar b b)$ one has
\begin{equation}
Z_{\rm sys} = 0.90\pm 0.03\;.
\label{6}
\end{equation}
The state density $dn/dM_n$ for the heavy quarkonium possesses the
empirical regularity, since with a good accuracy the differences between 
the state masses in the $(\bar c c)$ and $(\bar b b)$ systems do not 
depend on the flavors of heavy quarks. This regularity has found the
most evident expression in the framework of potential models as the
statement on the independence of the average kinetic energy of heavy quarks on
the flavors \cite{6}, so that the corresponding equation for the density of
energy levels takes place \cite{4}
\begin{equation}
\frac{dM_n}{dn}= \frac{2T}{n}\;.
\label{7}
\end{equation}
From (\ref{7}) one has
\begin{equation}
T=\frac{M_2-M_1}{\ln 4}\;,
\label{8}
\end{equation}
where $M_n$ is the mass of $nS$-level, $M_n=(3M_{Vn}+M_{Pn})/4$,
$M_{V,P}$ are the masses of vector and pseudoscalar states, correspondingly.
From the data on the masses of charmonium and bottomonium it follows that
$$
T=415\pm 20\; {\rm MeV.}
$$
Taking into account (\ref{7}), one has considered 
the spectroscopy leading to the estimates of the quark masses \cite{7}
\begin{equation}
m_b= 4.63\pm 0.03\; {\rm GeV,}\;\;\; m_c=1.18\pm 0.07\; {\rm GeV.}
\label{9}
\end{equation}
The $m_b$ value is close to that of given by M.Voloshin \cite{3},
$m^*_b = m_b(\mu)- 0.56\alpha_s(\mu)\mu\approx 4.64$ GeV. As for the $c$-quark 
mass, $m_c$ obtained in the $1/m_c^2$-order, its value makes only a weak 
influence on the accuracy of the $\alpha_s(m_Z)$ determination from the data on
$(\bar b b)$.

Note, that with a high accuracy the value
$$
a_Q =\alpha_s\; H_V\; \biggl(\frac{2m_Q}{M_1}\biggr)^2 Z_{\rm sys}
$$
does not depend on the heavy quark flavor, which leads to the scaling relations
for the leptonic constants \cite{4}. In accordance with (\ref{1}),
the empirical value
$$
a_Q = \pi\; \frac{f_1^2}{M_1}\; \frac{\ln 2}{M_2-M_1}
$$
gives a possibility to determine $\alpha_s$. In this way, 
one has to account for
the leptonic width of $\Upsilon(1S)\to l^+l^-$ is determined by the
effective fine structure constant $\bar \alpha_{em}$ at the $M_\Upsilon$
scale \cite{3}, so that we suppose $f_1=700\pm 15$ MeV.

In the two-loop accuracy for the "running" $\alpha_s$ 
constant\footnote{The analogous consideration at the
one-loop accuracy for $\alpha_s$ is given in \cite{8}.} (see 
prescriptions for the $\Lambda_{QCD}$ dependence on the number of active 
flavors in \cite{1}), one finds
\begin{equation}
\alpha_s^{\overline{\rm MS}}(m_Z)=0.106\pm 0.005\;,
\label{10}
\end{equation}
so that $\Lambda^{(5)}=108\pm 35$ MeV.
The error in (\ref{10}) is basically systematic. The increase of the
$b$-quark mass leads to the decrease of the mean value in (\ref{10}).
The uncertainty is determined by the variation of $m_b$ in the broad
region: $4.50< m_b<4.85$ GeV. The restriction of the $b$-quark mass,
as it stands in (\ref{9}), results in the decrease of the systematic error
from $0.005$ to $0.002$ in (\ref{10}).

The result of the $\alpha_s(m_Z)$ extraction in the framework of the
nonrelativistic sum rules for $(\bar b b)$ in the scheme of the spectral 
density moments \cite{3} supposes to have  a systematic uncertainty
close to 0.004. So, combining  (\ref{10}) with the result of \cite{3},
one gets the average value of coupling constant determined in the sum 
rules for $(\bar b b)$
\begin{equation}
\alpha_s^{\overline{\rm MS}}(m_Z)=0.108\pm 0.003\;.
\label{11}
\end{equation}
Note, that the recent application of the BLM scale choice to the hadronic 
event shape method of the $\alpha_s$ measurement resulted in 
$\alpha_s^{\overline{\rm MS}}(m_Z)=0.109\pm 0.008$ \cite{bur}.

The high accuracy of the $\alpha_s$ value extraction in the sum rules
for $(\bar b b)$ is caused by the large role of coulomb 
$\alpha_s/v$-corrections, which result in the linear dependence of the
leptonic constant squared on the value in (\ref{2}) at the scale of average
momentum transfer between the quarks inside the heavy quarkonium. In the 
way under consideration, this scale is equal to
\begin{equation}
\mu_b = e^{-5/6}\sqrt{2Tm_b} = 0.85\pm 0.03\; {\rm GeV.}
\label{12}
\end{equation}
In the Voloshin's paper \cite{3} $\mu_b=1$ GeV. From our point of view,
the difference between the $\mu_b$ values does mainly lead to the difference
between the $\alpha_s$ estimates given in these sum rules (\cite{3} and the
present work). The value inverse to (\ref{12}) determines the average size of 
the $(\bar b b)$ system. In the potential models it is close to 0.25 fm.

Finally, the analogous procedure for the $(\bar c c)$ system gives the
$\alpha_s$ value being in a good agreement with (\ref{10}), but the
corresponding uncertainty is much greater, because of the errors in the 
evaluation of $m_c$ and $Z_{\rm sys}$. On the other hand, the strong 
dependence of the $a_c$ value on the charm quark mass and the $\alpha_s$
extraction from the $(\bar b b)$ data can be used to evaluate $m_c$.
One finds
$$
m_c=1.20\pm 0.07\;\; {\rm GeV.}
$$
The uncertainty decreases to $0.02$ GeV if one uses the $b$-quark mass
in the strict region of (\ref{9}).

\section{Conclusion}

Thus, we have shown that the use of different schemes of sum rules for the
leptonic constants of vector states in the heavy $(\bar b b)$-quarkonium
leads to the systematically stable result 
$\alpha_s^{\overline{\rm MS}}(m_Z)=0.108\pm 0.003$, which is significantly lower
than the $\alpha_s(m_Z)$ value measured in the $Z$-boson pole at LEP.

The author expresses his gratitude to Academician S.S.~Gershtein and prof.
A.K.~Li\-kho\-ded for fruitful discussions and supports.
The author thanks also
O.P.~Yush\-chen\-ko (DELPHI) for private communications and a help.

This work is in part supported by the Russian Foundation for Basic Researches,
grants 96-02-18216, 96-02-04392 and by the Russian State Stipends for young 
scientists.

\vspace*{4mm}
\hfill {\it Received October 23, 1996}
\end{document}